\documentclass[twocolumn,english,final,showpacs,pra,lengthcheck,superscriptaddress]{revtex4-1}
\usepackage{ae,aecompl}
\usepackage[T1]{fontenc}
\usepackage[latin9]{inputenc}
\usepackage{color}
\usepackage{babel}
\usepackage{textcomp}
\usepackage{amsmath}
\usepackage{amssymb}
\usepackage{graphicx}
\usepackage[unicode=true,pdfusetitle,
 bookmarks=true,bookmarksnumbered=false,bookmarksopen=false,
 breaklinks=false,pdfborder={0 0 1},backref=false,colorlinks=true]
 {hyperref}

\makeatletter
\usepackage{babel}
\usepackage{amsfonts}

\providecommand{\U}[1]{\protect\rule{.1in}{.1in}}
\@ifundefined{textcolor}{}{
\definecolor{BLACK}{gray}{0}
\definecolor{WHITE}{gray}{1}
\definecolor{RED}{rgb}{1,0,0}
\definecolor{GREEN}{rgb}{0,1,0}
\definecolor{BLUE}{rgb}{0,0,1}
\definecolor{CYAN}{cmyk}{1,0,0,0}
\definecolor{MAGENTA}{cmyk}{0,1,0,0}
\definecolor{YELLOW}{cmyk}{0,0,1,0}
}

\usepackage{babel}

\makeatother

\begin{document}
\title{Manifestation of critical effects in environmental parameter estimation
using a quantum sensor under dynamical control}
\author{M. Cristina Rodr\'iguez}
\altaffiliation{maria.rodriguez@ib.edu.ar}

\affiliation{Centro At\'omico Bariloche, CONICET, CNEA, S. C. de Bariloche, Argentina.}
\affiliation{Instituto Balseiro, CNEA, Universidad Nacional de Cuyo, S. C. de Bariloche,
Argentina.}
\author{Analia Zwick}
\altaffiliation{analia.zwick@conicet.gov.ar}

\affiliation{Centro At\'omico Bariloche, CONICET, CNEA, S. C. de Bariloche, Argentina.}
\affiliation{Instituto Balseiro, CNEA, Universidad Nacional de Cuyo, S. C. de Bariloche,
Argentina.}
\affiliation{Instituto de Nanociencia y Nanotecnologia, CNEA, CONICET, S. C. de
Bariloche, Argentina}
\author{Gonzalo A. \'Alvarez}
\altaffiliation{gonzalo.alvarez@conicet.gov.ar}

\affiliation{Centro At\'omico Bariloche, CONICET, CNEA, S. C. de Bariloche, Argentina.}
\affiliation{Instituto Balseiro, CNEA, Universidad Nacional de Cuyo, S. C. de Bariloche,
Argentina.}
\affiliation{Instituto de Nanociencia y Nanotecnologia, CNEA, CONICET, S. C. de
Bariloche, Argentina}
\begin{abstract}
Quantum probes offer a powerful platform for exploring environmental dynamics, particularly through their sensitivity to decoherence processes. In this work, we investigate the emergence of critical behavior in the estimation of the environmental memory time $\tau_c$, modeled as an Ornstein-Uhlenbeck process characterized by a Lorentzian spectral density. Using dynamically controlled qubit-based sensors---realized experimentally via solid-state Nuclear Magnetic Resonance (NMR) and supported by numerical simulations---we implement tailored filter functions to interrogate the environmental noise spectrum and extract $\tau_c$ from its spectral width. Our results reveal a sharp transition in estimation performance between short-memory (SM) and long-memory (LM) regimes, reflected in a non-monotonic estimation error that resembles a phase transition. This behavior is accompanied by an avoided-crossing-like structure in the estimated parameter space, indicative of two competing solutions near the critical point. These features underscore the interplay between control, decoherence, and inference in open quantum systems. Beyond their fundamental significance, these critical phenomena offer a practical diagnostic tool for identifying dynamical regimes and optimizing quantum sensing protocols. By exploiting this criticality, our findings pave the way for adaptive control strategies aimed at enhancing precision in quantum parameter estimation---particularly in complex or structured environments such as spin networks, diffusive media, and quantum materials.
\end{abstract}
\maketitle

\section{\textit{\emph{Introduction}}}
Quantum technologies have enabled new paradigms in sensing, imaging, and metrology by exploiting the unique coherence properties of quantum systems. Among these, quantum sensing has emerged as a powerful approach in which quantum probes are used to extract environmental information---with high sensitivity and spatial resolution---from biological media, condensed matter systems, and spin environments \citep{QuantumSensing,pirandola_advances_2018,aslam_quantum_2023,demille_quantum_2024,bass_quantum_2024,deutsch2020harnessing}. These sensing protocols have been implemented across a range of platforms, including superconducting qubits \citep{bylander2011noise, von2020two}, trapped ions \citep{frey2020simultaneous}, nitrogen-vacancy (NV) centers in diamond \citep{Staudacher2013,Shi2013,staudacher_probing_2015,romach2015spectroscopy,schmitt2017submillihertz,hernandez2024optimal,Rosenberg2025}, and nuclear spins \citep{Alvarez2011,Alvarez2013,Smith2012,alvarez_controlling_2012,zwick2020,yang_probe_2020,zhang_detection_2008,jiang_multiparameter_2021}. These platforms enable precise monitoring of decoherence---the loss of quantum coherence due to interactions with the environment---which provides a direct window into the environment's dynamical properties \citep{Zurek2003,suter2016colloquium,szankowski2017environmental,Zwick2023}.

A key environmental feature accessible via decoherence monitoring is the memory (or correlation) time $\tau_c$, which characterizes how long environmental fluctuations persist \citep{Alvarez2013,Alvarez2013_JMR,zwick2016maximizing,zwick2020,Zwick2023}. Estimating $\tau_c$ is central to applications across diverse fields: it enables characterization of biological systems through molecular diffusion \citep{Alvarez2013,Alvarez2013_JMR,Shemesh2015,zwick2016maximizing,zwick2020,Capiglioni2021}; identification of chemical-exchange processes \citep{Smith2012}; exploration of quantum phase transitions in spin environments \citep{Haikka2012,Gessner2014}; probing of non-local correlations in composite systems \citep{Smirne2013,Laine2012,Liu2013,Wissmann2014}; and the study of charge and spin diffusion in materials and spin networks \citep{Feintuch2004,Suter1985,Alvarez2011,Alvarez2013a,alvarez_controlling_2012,Alvarez2015,Dominguez2021a,Dominguez2021,Kuffer2024,Rosenberg2025}.

Many protocols designed to characterize parameters from the environment use dynamical-decoupling techniques \citep{viola1998dynamical,vio99,kofman_universal_2001,kofman_unified_2004,Khodjasteh2005,gordon_Universal_2007,gordon_optimal_2008,souza_robust_2012,suter2016colloquium}. These techniques are based on the ability of the dynamical-decoupling sequences to filter the noise \citep{Alvarez2011,Bylander2011,yuge2011measurement}. Studying how the applied control affects the estimation of the desired parameters is a crucial task to find the optimal strategy to obtain information about the environment \citep{zwick2016maximizing,zwick2020,Zwick2023,ronchi_maximizing_2024,Kuffer2024}.

Of particular interest are environmental processes modeled as Ornstein-Uhlenbeck (OU) noise, which describes dynamics such as molecular diffusion in biological tissues \citep{Alvarez2013,Alvarez2013_JMR,Shemesh2015,zwick2016maximizing,zwick2020,Capiglioni2021}, charge transport in solids \citep{Feintuch2004}, and spin diffusion in interacting spin networks \citep{Suter1985,Alvarez2011,Alvarez2013a,alvarez_controlling_2012,Alvarez2015,Dominguez2021a,Kuffer2024,Rosenberg2025}. These environments induce an exponentially decaying autocorrelation function $\langle B(t)B(0)\rangle \propto e^{-t/\tau_c}$, where $\tau_c$ is the correlation time and $B(t)$ represents the stochastic field acting on the quantum sensor. In the frequency domain, this behavior leads to a Lorentzian spectral density 
\begin{equation}
G(\tau_{c},\omega)\propto\frac{\tau_{c}}{1+\omega^{2}\tau_{c}^{2}}.\label{eq:G-1}
\end{equation}
whose width encodes $\tau_c$, making it a key parameter in characterizing the environment.

The precision of estimating this environmental memory time parameter
may exhibit critical behavior as a function of the memory time \citep{zwick2016criticality}.
This criticality manifests as a divergence in the lower bound of the
estimation error $\varepsilon_{R}=\frac{\delta\tau_{c}}{\tau_{c}}$,
as derived from quantum information theory, but only when combined
with proper dynamical control. Analogous to a phase transition, this
behavior defines a sharp boundary between the short- and long-memory
regimes in the probe\textquoteright s decoherence dynamics, corresponding
to short- and long-memory of the environment relative to the qubit-probe
measurement time, respectively. In contrast, the free evolution of
the probe---corresponding to free induction decay (FID)---exhibits
a smooth transition between these two dynamical regimes \citep{zwick2016criticality},
reflecting the gradual shift from non-Markovian to Markovian behavior
\citep{Breuer_RevModPhys.88.021002,Rivas2014}.

The uncovered criticality similar to a phase transition in the memory-time information
represents a fundamental characteristic that defines dynamical behavior
and plays a crucial role in achieving optimal precision for estimating
the environment\textquoteright s memory time within experimental constraints.
Paradoxically, while the sharp crossover corresponds to an absence
of information at the critical point, it also serves as a distinctive
signature of dynamical behavior. This duality has profound practical
implications: It not only enables maximum estimation precision for
the memory time, but also guides the development of tailored control
strategies to optimize information extraction \citep{zwick2016criticality}.

Although the theoretical prediction of such criticality is established
\citep{zwick2016criticality}, its experimental manifestation remains
intriguing, as it relies on the application of proper dynamical control
to the quantum probe. The critical point and optimal control regimes
to extract the maximum information about the memory time explicitly
depend on its unknown value \citep{zwick2016criticality}, which poses
a significant challenge for experimental implementation where the
correlation time is unknown. To address this, it is essential to understand
how this criticality emerges in practice.

In this work, we address the challenge of environmental parameter estimation by implementing solid-state Nuclear Magnetic Resonance (NMR) experiments on a system that exhibits theoretically predicted information-theoretic critical behavior. Through a combination of experimental data and numerical simulations, we benchmark our results against analytical predictions, focusing specifically on the estimation of the environment's memory (correlation) time $\tau_c$. We observe the emergence of two competing solutions for $\tau_c$ as a function of the control or observation time, corresponding to distinct dynamical regimes of the sensor's evolution---only one of which yields accurate estimations in each regime.

A sharp transition between these solutions reveals a critical crossover resembling an avoided level crossing, marking a boundary between the short-memory and long-memory regimes. This criticality not only enables regime identification but also guides the design of optimized quantum sensing strategies. In particular, it suggests that tuning control protocols near the critical point can enhance sensitivity and reduce the number of required measurements. 

Moreover, the ability to dynamically control the quantum sensor enables the design of tailored filter functions that selectively probe different regions of the environmental spectrum. This opens avenues for spectral engineering and targeted noise sensing, allowing quantum probes to extract relevant information from environments with structured or broadband noise characteristics. This approach might also be relevant in materials where spin or charge dynamics give rise to correlated fluctuations, such as quantum magnets, low-dimensional conductors, or disordered spin networks---systems of growing interest for quantum sensing applications.

\section{Controlled qubit-probe as a sensor of the environment memory time
and critical dynamical behaviors}
\label{0}\foreignlanguage{american}{}
\begin{figure}
\centering{}
\includegraphics[width=1\columnwidth]{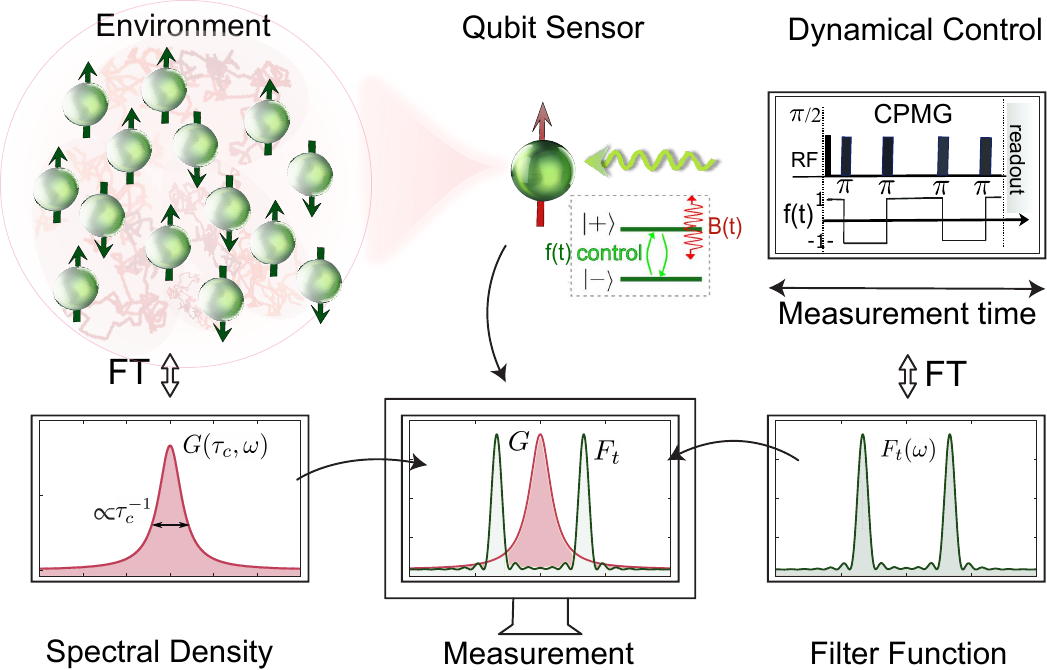}
\caption{\label{fig1_scheme}A dynamically controlled qubit probe senses the
environmental memory (correlation) time $\tau_{c}$ while undergoing
pure dephasing due to its interaction with the environment. The environment
induces an effective fluctuating field $B(t)$ on the qubit sensor,
whose influence is modulated by dynamical control with $f(t)$. This modulation results in an attenuation factor in the measured magnetization signal of the qubit probe. It is determined by the overlap between the control filter function $F_{t}(\omega)$ and the environmental spectral density $G(\tau_{c},\omega)$, which are the Fourier transforms (FT) of the control modulation $f(t)$ and the environmental correlation function $\left\langle B(t)B(0)\right\rangle$, respectively. Assuming an Ornstein-Uhlenbeck process, i.e., $\left\langle B(t)B(0)\right\rangle \propto e^{-t/\tau_{c}}$, the spectral density $G(\tau_c, \omega)$ takes a Lorentzian shape with a width proportional to $\tau_c^{-1}$, the inverse of the memory time. The dynamical control illustrated here is the Carr-Purcell-Meiboom-Gill (CPMG) sequence, which consists of equidistant $\pi$-pulses applied after an initial RF $\pi/2$-pulse. In terms of $f(t)$, this corresponds to flipping the sign ($\pm1$) of the interaction coupling between the sensor and its environment. This standard decoupling sequence repeatedly swaps the qubit state between $|+\rangle$ and $|-\rangle$. The filter function $F_{t}(\omega)$ for the CPMG sequence consists mainly of two symmetric sinc functions centered at the fundamental frequency $\pm \omega_{\text{ctrl}}$. As the number of pulses $N$ increases, $F_{t}(\omega)$ becomes a narrower bandpass filter, approaching a Dirac delta function.}.
\end{figure}

We consider a dynamically controlled qubit-probe experiencing pure
dephasing due to the probe-environment interaction $\sigma_{z}gB$,
where $g$ represents the coupling strength, $\sigma_{i}$ are the
Pauli operator of the qubit, and $B$ is the environment operator
(Fig. \ref{fig1_scheme}). In the weak-coupling regime, this decoherence
effect is characterized by an attenuation factor $\mathcal{J}(\tau_{c},t)$,
which decreases the observable qubit-magnetization as a function of
the evolution time

\begin{equation}
\langle\sigma_{x}(t)\rangle=\langle\sigma_{x}(0)\rangle\,e^{-\mathcal{J}(\tau_{c},t)}.\label{eq:Magnertization}
\end{equation}

\noindent Here $\langle\sigma_{x}(0)\rangle$ denotes the initial
state of the qubit-probe, $t$ is the total evolution/control time
and $\tau_{c}$ is the memory (correlation) time of the environment.
The information about $\tau_{c}$ is encoded in the attenuation factor
$\mathcal{J}(\tau_{c},t)$ which also depends on the overlap between
a control filter function $F_{t}(\omega)$ and the environment spectral
density $G(\tau_{c},\omega)$ \citep{kofman_universal_2001,Zwick_ChapGK_2015,zwick2016criticality,zwick2016maximizing,Kuffer_PRXQuantum.3.020321}

\begin{equation}
\mathcal{J}(\tau_{c},t)=\int_{-\infty}^{\infty}F_{t}(\omega)\,G(\tau_{c},\omega)d\omega.\label{eq:attenuation_factor}
\end{equation}

\noindent The control filter function $F_{t}(\omega)$ is proportional
to the square of the Fourier transform of the modulation function $f(t)$
due to dynamical control applied on the qubit during the control/measurement
time $t$, while $G(\tau_{c},\omega)$ is the Fourier transform of
the correlation function of the environment. Here we explicitly remark
the dependency on $\tau_{c}$ because it is the parameter of interest
in this work, however $G$ is characterized by all the parameters
that describe the environment (Fig. \ref{fig1_scheme}). In this work,
we consider a spectral density with a Lorentzian shape representing
Ornstein-Uhlenbeck processes (Eq. (\ref{eq:G-1})).

We demonstrated that the estimation precision of $\tau_{c}$ can display
critical behavior depending on the memory-time parameter \citep{zwick2016criticality}.
This criticality is observed in the divergency of the error in the
estimation of $\tau_{c}$ resembling a phase transition, is an inherent
property of the environmental noise spectrum (Eq. (\ref{eq:G-1})),
and becomes apparent only under proper dynamic control. It establishes
a distinct boundary between the probe\textquoteright s short-memory
and long-memory decoherence regimes relative to the environmental
memory time, i.e., $t>\tau_{c}$ and $t<\tau_{c}$, where $t$ represents
the probing time.

To manifest this criticality, the control on the qubit probe should generate a very narrow bandpass filter \( F_t(\omega) \), such that it approximates a Dirac delta function centered at the control frequency \( \omega_{\text{ctrl}} \), i.e., \( F_t(\omega) \propto \delta(\omega - \omega_{\text{ctrl}}) \). In this case, the value \( F_t(\omega_{\text{ctrl}}) \) effectively determines the weight of the environmental spectral density in the integration. In the next Section, we discuss which kind of control produces such
filter. Then from Eq. (\ref{eq:attenuation_factor}), the
attenuation factor becomes $\mathcal{J}(\tau_{c},t)\propto F_{t}(\omega_{ctrl})\,G(\tau_{c},\omega_{ctrl})$
and its derivative with respect to the parameter to be estimated $\tau_{c}$
is null at $\omega_{ctrl}\approx\tau_{c}^{-1}$
\begin{equation}
\left.\frac{\partial\mathcal{J}}{\partial\tau_{c}}\right|_{\omega_{ctrl}\approx\tau_{c}^{-1}}\propto\left.\frac{\partial G}{\partial\tau_{c}}\right|_{\omega_{ctrl}\approx\tau_{c}^{-1}}\approx0.
\end{equation}
This makes the Quantum Fisher information $\mathcal{F_{Q}}$ \citep{Caves_1994_fisher,Paris2009_QUANTUM-ESTIMATION}
null at $\omega_{ctrl}=\tau_{c}^{-1}$ \citep{zwick2016criticality},
which induces a divergence of the minimum achievable relative-error
(per measurement) $\varepsilon_{\mathcal{F}}$ of the (unbiased) estimation
of $\tau_{c}$, $\varepsilon_{R}=\frac{\delta\tau_{c}}{\tau_{c}}$,
as predicted by the Cramer-Rao bound limit
\begin{equation}
\varepsilon_{R}\geq\varepsilon_{\mathcal{F}}(\tau_{c},t)=\frac{1}{\tau_{c}\sqrt{\mathcal{F_{Q}}(\tau_{c},t)}},\,\,\mathcal{F_{Q}}=\frac{e^{-2\mathcal{J}}}{1-e^{-2\mathcal{J}}}\left(\frac{\partial\mathcal{J}}{\partial\tau_{c}}\right)^{2}.\label{eq:error_QFI}
\end{equation}

\begin{figure}[t]
\centering{}
\includegraphics[width=1\columnwidth]{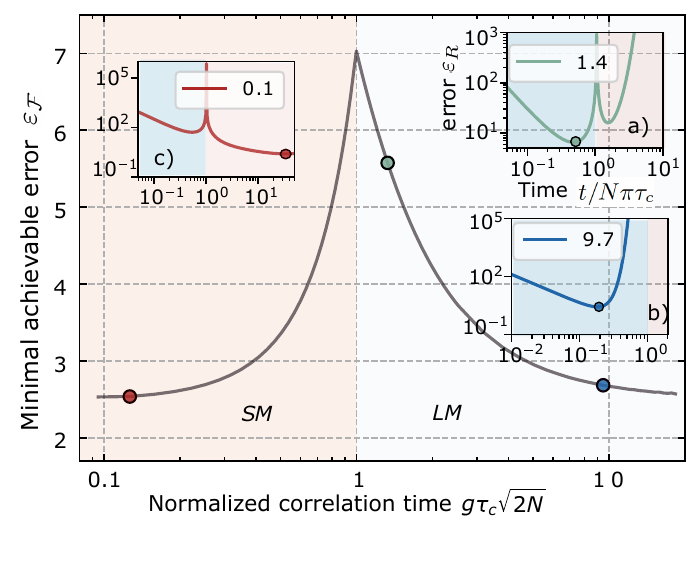}\caption{Critical transition in the minimum achievable relative error $\varepsilon_{\mathcal{F}}$ per measurement
(Eq. (\ref{eq:error_QFI})) for estimating the correlation time $\tau_{c}$
as a function of the environmental parameters ($\tau_{c}$ and interaction
strength $g$) and the number of pulses $N$ in the control sequence
(such as CPMG and CW). When $g\tau_{c}\sqrt{2N}>1$ the minimum error $\varepsilon_{\mathcal{F}}$
occurs in the LM regime, whereas for $g\tau_{c}\sqrt{2N}<1$, the
minimum $\varepsilon_{\mathcal{F}}$  corresponds to the SM regime. Insets: Relative error $\varepsilon_{R}=\frac{\delta\tau_{c}}{\tau_{c}}$
plotted as a function of the normalized time $t/N\pi\tau_{c}$ for
specific parameter sets: (a) $g\,\tau_{c}\,\sqrt{2N}=1.4$; (b) $g\,\tau_{c}\,\sqrt{2N}=9.7$
and (c) $g\,\tau_{c}\,\sqrt{2N}=0.13$. The relative error $\varepsilon_{R}$
have two minimum values and diverge around $t\thickapprox N\pi\tau_{c}$.
The global minimum achievable error in each inset, indicated by a colored dot, defines  $\varepsilon_{\mathcal{F}}$. This point is then shown in the main plot using the corresponding color from each inset. In (a) and (b), the dot corresponds to the LM regime (highlighted in blue), while in (c), it corresponds to the SM regime (highlighted in red). These plots illustrate the transition between regimes in estimation precision}
\label{fig2_Criticality-schematic}
\end{figure}
This divergence arises from the intrinsic nature of the spectral density in Eq. (\ref{eq:G-1}), whose derivative \( \frac{\partial G}{\partial \tau_c}  \) is zero at \( \omega = \tau_c^{-1} \).
This point that gives no information about $\tau_{c}$ indicates a
sharp transition between two dynamical regimes for the attenuation-factor:
the short-memory (SM) $\mathcal{J}^{SM}$ for $t\gg\tau_{c}$ and
long-memory (LM) $\mathcal{J}^{LM}$ for $t\ll\tau_{c}$. Each regime
offers a local minimum estimation error dictated by the Cramer-Rao
bound limit, with one being the global minimum (Fig. \ref{fig2_Criticality-schematic}).
The global minimum indicates the dynamical control strategy to follow
to achieve the best precision in the estimation of $\tau_{c}$ \citep{zwick2016criticality}.

The critical point at $\omega_{ctrl}=\tau_{c}^{-1}$ and the optimal
control regimes (where to extract the maximum information of the memory
time) explicitly depend on the unknown value of $\tau_{c}$ \citep{zwick2016criticality}.
This anticipates a difficulty for experimental implementation where
naturally the parameter to be estimated is unknown. While there are
strategies to bypass this limitation by real-time estimation processes
\citep{zwick2016maximizing}, this critical behavior and how it will affect the estimation process of the parameter have yet to be explored. Here
we thus study how the critical behavior is manifested during an experimental
estimation process contrasted by numerical simulations, and the theoretical
predictions. 

\section{Estimation of the environmental memory time\label{sec:III_Estimation-of-the}}

We estimate the correlation time by extracting its value
from a measured qubit-probe magnetization signal. This signal provides
the attenuation factor (Eq. (\ref{eq:Magnertization})), from which
the $\tau_{c}$ can be inferred. Here, we derive the analytical expression
for the attenuation factors and the corresponding formula for estimating
$\tau_{c}$.

The criticality in the estimation error of $\tau_{c}$ is evidenced
under a control producing a narrow bandpass filter \citep{zwick2016criticality}.
Such spectral filter can be achieved by dynamically controlling the
qubit-probe with a sufficient large number $N$ of continue-wave (CW)
oscillations or equidistant $\pi$-pulses over the total control time $t$. The decoupling
sequence of equidistant $\pi$-pulses is known as the Carr-Purcell-Meiboom-Gill (CPMG) sequence \foreignlanguage{american}{\citep{Carr1954,Meiboom1958,Slichter1990}},
Fig. \ref{fig1_scheme}.  Under such
control, the filter function $F_{t}(\omega)$, as well the attenuation
factor $\mathcal{J}(\tau_{c},t)$ {[}Eq. (\ref{eq:attenuation_factor}){]}
can be analytically derived \citep{ajoy2011optimal,zwick2016criticality,zwick2016maximizing}.

For $N\gg1$, $F_{t}(\omega)$ converges to a sum of narrow sinc functions
centered at the harmonics of the inverse modulation-period $k\omega_{ctrl}=\pi Nk/t$,
$k\in\mathbb{Z}$ \citep{ajoy2011optimal}. In this narrow-multi-band
pass-filter (NF) approximation, the attenuation factor becomes $\mathcal{J}(\tau_{c},t)\approx\sum_{k=1}^{\infty}F_{t}(k\omega_{ctrl})\,G(\tau_{c},k\omega_{ctrl})$,
where the dominant contribution arises from the first harmonic frequency
$\omega_{ctrl}$ (corresponding to $k=1$), being indeed the only
contribution under CW, while the effects of higher-order harmonics
can be neglected for CPMG \citep{ajoy2011optimal}.

In the following, we consider the CPMG sequence control which is the control applied to the experiments. Under
this approximation, the attenuation factor is
\begin{equation}
\mathcal{J}^{NF}(\tau_{c},t)=F_{t}(\omega_{ctrl})\,G(\tau_{c},\omega_{ctrl})=\frac{g\text{\texttwosuperior}\tau_{c}t}{1+(\omega_{ctrl}\tau_{c})^{2}},\label{eq:J1harmonic}
\end{equation}
and an analytical expression can be derived to extract the $\tau_{c}$
value from the signal attenuation factor

\begin{equation}
\tau_{c}^{NF}(t)=\frac{g^{2}t^{3}}{2\pi^{2}N^{2}\mathcal{J}^{NF}(t)}\left[1\pm\sqrt{1-\left(\frac{2\pi^{3} N \mathcal{J}^{NF}(t)}{g^{2}t^{2}}\right)^{2}}\right],\label{eq:tauc_1_NF}
\end{equation}
where two possible solutions for $\tau_{c}$ exist. We can approximate
these solutions when the probing time is in the short- and long-memory
regimes.

Under CPMG control, when the probing time between pulses is shorter
than the memory time $t/N\ll\tau_{c}$ corresponds to the LM regime,
which is strongly dependent on the spectral density\textquoteright s
high-frequency components. Conversely, in the long-time limit, $t/N\gg\tau_{c}$,
corresponds to the SM regime, where the filter function is broader
than the spectral density. The attenuation factors in these regimes
are approximately \citep{zwick2016criticality}\foreignlanguage{american}{
\begin{eqnarray}
\mathcal{J}^{LM}\!\! & \simeq & \!\frac{g^{2}t^{3}}{12\!N^{2}\tau_{c}},\mathcal{\:J}^{SM}\!\!\simeq\!g^{2}\tau_{c}t.\label{eq:J}
\end{eqnarray}
}

\noindent Then the analytical expressions for extracting $\tau_{c}$
on these limits are

\begin{equation}
\tau_{c}^{SM}(t)=\frac{\mathcal{J}^{SM}(t)}{tg^{2}},\tau_{c}^{LM}(t)=\frac{g^{2}t^{3}}{12\!N^{2}\mathcal{J}^{LM}(t)}.\label{eqtaucSM_LM}
\end{equation}
We can see that one of the solutions of Eq. (\ref{eq:tauc_1_NF})
corresponds to the expected true value of $\tau_{c}$ within the LM-regime
$t<N\pi\tau_{c}$, and the other for SM-regime $t>N\pi\tau_{c}$.

The attenuation factor $\mathcal{J}$ under CPMG control has analytical
solution \citep{ajoy2011optimal,Alvarez2011,Alvarez2013}. Thus we
also consider this exact expression to estimate the correlation time.
Although the analytical expression cannot be inverted analytically
to extract $\tau_{c}$, it can be numerically solved. Thus, for
each measurement time, we numerically obtain two possible values of $\tau_{c}$.

\section{Estimating the memory-time from simulated experiments\label{sec:Estimating-the-memory-time}}

To understand the estimation process near the regime of critical behavior, we first simulate experimental data. This helps identify how critical behavior affects the estimation of the correlation time.
Specifically, we numerically
generate the qubit signal under a CPMG control sequence, where the
qubit interacts with an environment characterized by a Lorentzian
spectral density. This NMR control sequence and the spectral density
parameters are consistent with the experimental analysis introduced in the next section
to study the behavior of the estimation process.

To analyze the dynamics across different controlled regimes, we vary
the environmental parameters --- the probe-environment
interaction strength $g$ and the correlation time $\tau_{c}$ --- as
well as the control parameters, including the number of pulses $N$
and the measurement time $t$. We consider three representative cases,
characterized by the dimensionless parameter $g\,\tau_{c}\,\sqrt{2N}=0.13,1.4$
and $9.7$, as illustrated in Fig. \ref{fig2_Criticality-schematic}.
Each case corresponds to a different optimal estimation regime for
$\tau_{c}$, based on theoretical predictions. In the case of $g\,\tau_{c}\,\sqrt{2N}=0.13$,
the short-memory (SM) regime provides the best estimation of $\tau_{c}$,
while for $g\,\tau_{c}\,\sqrt{2N}=9.7$, the long-memory (LM) regime
is more favorable. The intermediate case, $g\,\tau_{c}\,\sqrt{2N}=1.4$,
lies at the transition between these regimes. Although the LM regime
is optimal here, both regimes exhibit comparable estimation efficiency,
making it an ideal scenario to observe the critical behavior in the
information about $\tau_{c}$.

\begin{figure*}[!t]
\centering{}

\includegraphics[width=1\textwidth]{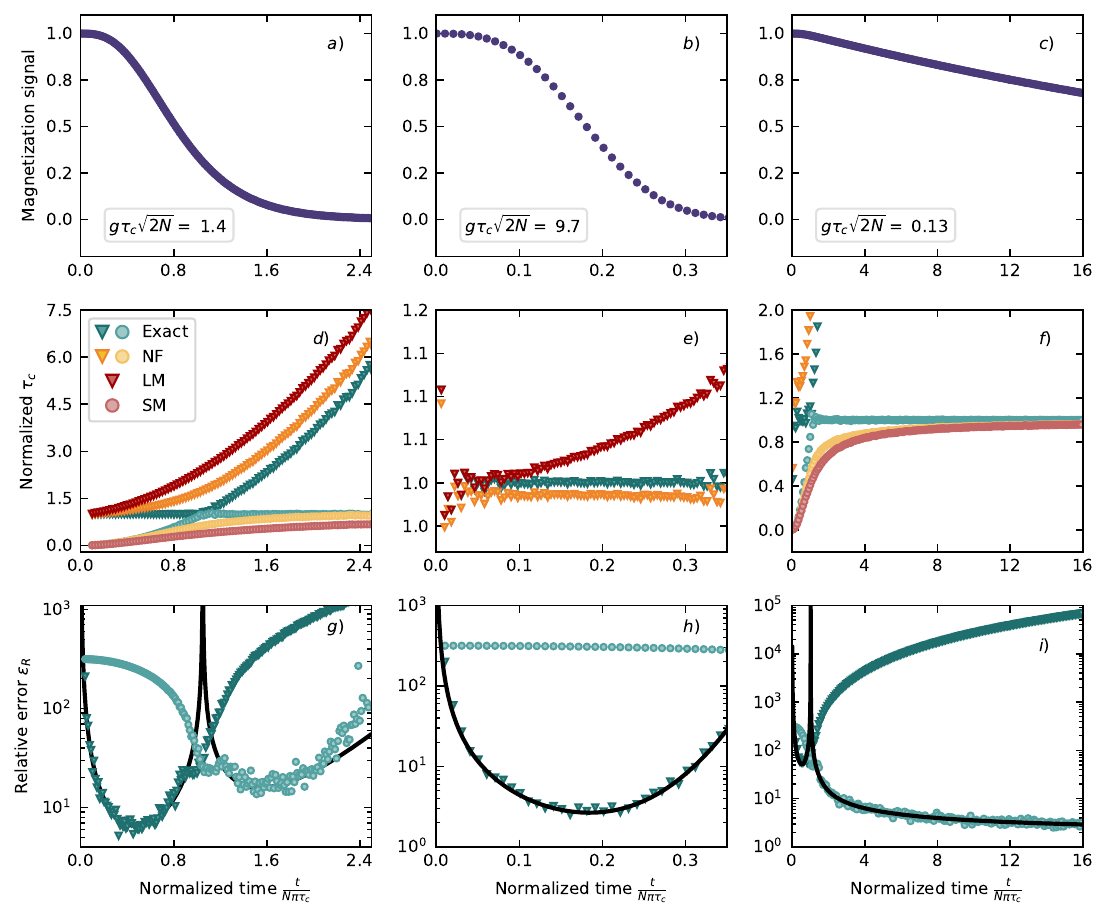}\caption{Simulation of a qubit probe under CPMG control with $N$ pulses, interacting
with an environment characterized by a Lorentzian spectral density.
The selected parameters correspond to the three dynamical regimes
indicated in Fig. \ref{fig2_Criticality-schematic}: near the critical
transition ($g\tau_{c}\sqrt{2N}=1.4$), the long-memory (LM) regime
($g\tau_{c}\sqrt{2N}=9.7$), and the short-memory (SM) regime ($g\tau_{c}\sqrt{2N}=0.13$).
(a)--(c) Magnetization signal as a function of the renormalized time
$t/N\pi\tau_{c}$. (d)--(f) Normalized estimated memory (correlation)
time $\tau_{c}/\tau_{c}^{\text{true}}$ as a function of time, obtained
using different estimation approaches: the exact attenuation factor
$\mathcal{J}$, the first harmonic approximation $\mathcal{J}^{NF}$,
the SM regime approximation $\mathcal{J}^{SM}$, and the LM regime
approximation $\mathcal{J}^{LM}$. The exact attenuation factor $\mathcal{J}$,
as well as $\mathcal{J}^{NF}$ provide two possible solutions for
$\tau_{c}$, represented by green and orange (dark triangle and light
circle) respectively. In contrast, the approximations $\mathcal{J}^{SM}$
and $\mathcal{J}^{LM}$ yield unique values representative of their
respective regimes, shown in red (dark triangle and light circle).
As expected, the estimation is accurate ($\tau_{c}\approx\tau_{c}^{\text{true}}$)
within the validity range of each regime. In (d) and (f), the critical
transition manifests as an avoided crossing between solutions at $t\sim N\pi\tau_{c}$
under the $\mathcal{J}^{NF}$ approximation, while for the exact $\mathcal{J}$,
it appears as a crossover of the correct, physically meaningful solution.
In (e), this regime is not reached due to signal numerical noise,
preventing the identification of the transition. (g)--(i) Relative
error per measurement $\varepsilon_{R}$ in the estimation of $\tau_{c}$
as a function of the renormalized time. The error is computed as the
ratio of the standard deviation to the true value $\tau_{c}^{\text{true}}$
across the two numerically obtained exact solutions (green dark triangle
and light circle curves) and is compared with the theoretical Cramer-Rao
bound (black curve) derived from the quantum Fisher information (QFI)
in Eq. (\ref{eq:error_QFI}). The critical behavior is highlighted
by variations in the relative error near the transition between dynamical
regimes. This figure illustrates the interplay between estimation
approaches and dynamical regimes in determining $\tau_{c}$. For simulations
with $g\tau_{c}\sqrt{2N}\approx1.4,9.7$, we use the experimentally
estimated values $g=8.58$ kHz and $\tau_{c}=0.08$ ms (Sec. \ref{sec:Estimation-error-analysis})
with $N=2$ and $N=100$, respectively. The experimental values prevent
achieving $g\tau_{c}\sqrt{2N}\approx0.13$ by lowering $N$, so for
this case, we use $g=1$, $\tau_{c}=0.02$, and $N=20$.}
\label{Fig3}
\end{figure*}

This numerical approach simulates an experimental procedure where
the environmental parameters are unknown, closely mimicking real experimental
conditions. To simulate the experimental magnetization signal given
by Eq. (\ref{eq:Magnertization}) and shown in Fig. \ref{Fig3} (upper
rows), we generate spin state measurements based on the probability
of obtaining the $|+_{x}\rangle$ (spin up) or $|-_{x}\rangle$ (spin
down) state in a measurement of $\sigma_{x}$. This probability is
given by
\begin{equation}
p_{\pm}(\tau_{c},t)=\frac{1}{2}(1\pm e^{-\mathcal{J}(\tau_{c},t)}).\label{proba-1}
\end{equation}

For each measurement time, we assign a spin state by generating a
random number between 0 and 1 and comparing it with $p_{\pm}(\tau_{c},t)$.
If the random number falls within the probability range for $|+_{x}\rangle$,
the spin is assigned as up; otherwise, it is assigned as down. This
stochastic sampling mimics experimental quantum statistical fluctuations
inherent in real measurements.

To obtain a reliable estimate of the magnetization signal, we measure $N_{m}=10^5$ times for each measurement time,
averaging the results to compute the mean magnetization. This mimics an NMR experiment, where the signal corresponds to the average over an ensemble of independent spins acting as qubit sensors, each interacting with its own environment. Additionally,
this procedure is repeated $N_{\text{rep}}=50$ times to determine the final
mean magnetization value and its standard deviation, allowing us to
capture statistical variations across independent measurements. The
final averaged magnetization represents the simulated experimental
data shown in Fig. \ref{Fig3} (upper rows) as a function of the renormalized
time $t/N\pi\tau_{c}$. This approach ensures that the numerical simulations
not only reproduce the expected average behavior but also reflect
the fluctuations inherent in experimental measurements.

To support the experimental data presented in the next section, the
simulations in Fig. \ref{Fig3}(a)-(b) assume $g=8.58$ kHz and
$\tau_{c}=0.08$ ms as the true value, with $N=2$ and $N=100$, respectively, to represent
$g\,\tau_{c}\,\sqrt{2N}=1.4$ and $9.7$. In Fig. \ref{Fig3}(b),
the magnetization decays rapidly, leading to a noisy signal at longer
times beyond the displayed range. The experimental values of $g$
and $\tau_{c}$ prevent us from achieving $g\,\tau_{c}\,\sqrt{2N}\approx0.13$
by simply lowering $N$. Therefore, for the corresponding simulations
in Fig. \ref{Fig3}(c), we select a representative case with $g=1$,
$\tau_{c}=0.02$ ms as the true value, and $N=20$.

The estimation of $\tau_{c}$ as a function of time is shown in Fig.
\ref{Fig3} (middle rows) for the different estimation strategies
discussed in Sec. \ref{sec:III_Estimation-of-the}. The vertical axis
represents the normalized estimated correlation time relative to its
true value, $\tau_{c}/\tau_{c}^{\text{true}}$, where $\tau_{c}^{\text{true}}$
is the value used to simulate the magnetization signals in each case.
For the exact expression of $\mathcal{J}$, the two numerically estimated
values of $\tau_{c}$ are shown in dark and light green. In the SM
regime ($t>N\pi\tau_{c}$), the dark green estimation corresponds
to the true value and thus represents the physically meaningful solution.
Conversely, in the LM regime ($t<N\pi\tau_{c}$), the light green
estimation provides the correct value. Near $t\approx N\pi\tau_{c}$,
these solutions exhibit a crossing.

We now examine the approximate expressions derived for the narrow-filter
control, which predicts the observation of critical behavior in the estimation process. The narrow band-pass filter approximation remains
valid within the observed region for $t/N \pi \tau_c > 2$ and for $t/N \pi \tau_c < 0.5$. The two solutions for $\tau_{c}^{NF}$, estimated from
the narrow-band filter approximation $\mathcal{J}^{NF}$ in Eq. (\ref{eq:tauc_1_NF}),
are shown in dark and light orange. These solutions exhibit an avoided-crossing
behavior near the critical condition at $t\approx N\pi\tau_{c}$.
Notably, this estimation process provides the true value of $\tau_{c}$
only in the extreme conditions that define the LM and SM regimes,
i.e., $t/N \pi \tau_c < 0.5$ and $t/N \pi \tau_c > 2$. The avoided-crossing behavior
represents a fundamental aspect of the transition between dynamical
regimes and serves as a hallmark of criticality in the estimation
process.

A similar pattern is observed for the estimated correlation times
obtained using the LM and SM regime approximations, $\tau_{c}^{LM}$
and $\tau_{c}^{SM}$, as given in Eq. (\ref{eqtaucSM_LM}) and shown
in dark and light red. The vertical axis represents the
normalized estimated correlation time $\tau_{c}/\tau_{c}^{\text{true}}$,
where $\tau_{c}^{\text{true}}$ is the value used to simulate the
magnetization signals in each case. The different estimation approaches
yield distinct results depending on the dynamical regime. The LM approximation
is expected to be accurate in the regime where $t/N  \ll \tau_c$, while the SM approximation
should hold in the regime where $t/N \gg \tau_c$. However, near the transition, discrepancies
arise, highlighting the limitations of these approximations outside
their respective validity ranges.

.

\section{Estimating the memory-time from solid-state NMR experimental data\label{2}}

To validate the theoretical predictions discussed in previous sections,
we estimate the memory time $\tau_{c}$ from experimental data obtained
in a solid-state NMR system. We select an experimental case where
the noise spectral density closely resembles the Lorentzian model
used in our theoretical framework. This choice allows us to assess
the presence of the critical behavior predicted by our model and evaluate
the performance of different estimation strategies. We perform NMR
experiments on a polycrystalline adamantane sample, where the spin-$\frac{1}{2}$
nuclei of $^{13}C$ serve as quantum probes of the surrounding $^{1}H$
spin bath. Each $^{13}C$ spin is coupled to multiple $^{1}H$
spins through magnetic dipole interactions.

\begin{figure}[t]
\centering{}\includegraphics[width=0.97\columnwidth]{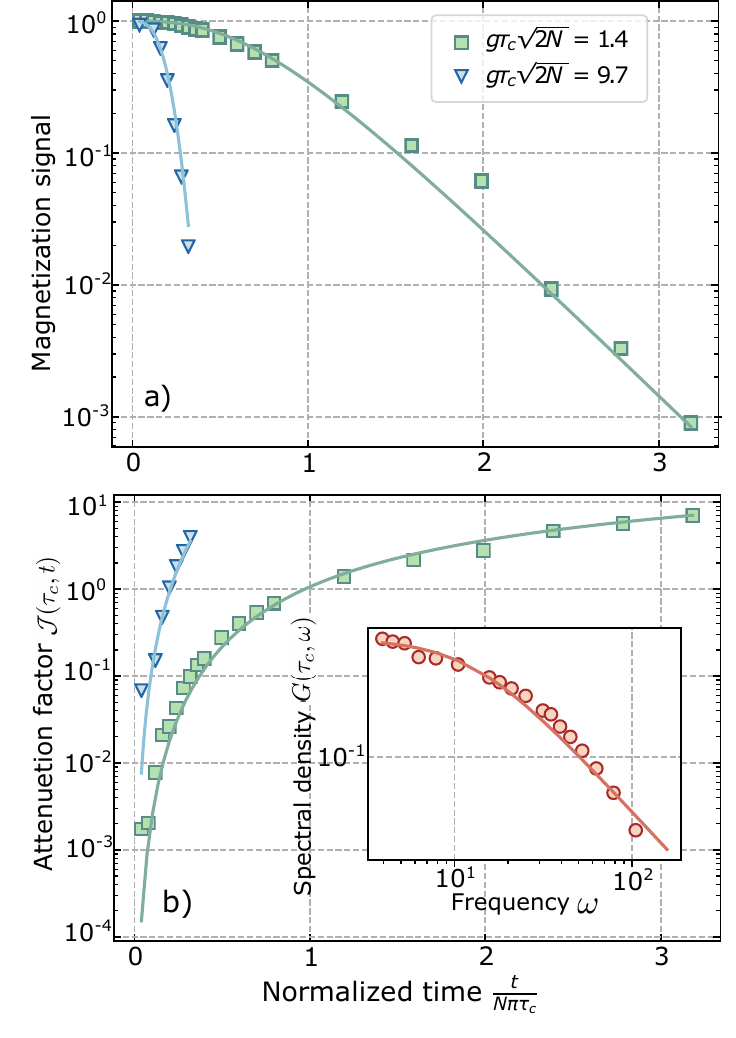}
\caption{Experimental measurement and analysis of the magnetization signal
using CPMG sequences with a fixed number of pulses ($N=2,100$) and
variable delays between pulses to adjust the total measurement time.
(a) The experimental magnetization signal as a function of time. The
spectral density is approximately Lorentzian (Eq.(\ref{eq:G-1}))
with environmental parameters $g=8.58\,\text{kHz}$ and $\tau_{c}=0.08\,\text{ms}$.
The corresponding dimensionless parameters are $g\tau_{c}\sqrt{2N}=1.4$
(for $N=2$, green squares) and $g\tau_{c}\sqrt{2N}=9.7$ (for $N=100$,
blue triangles). The qubit probes are the $^{13}C$ nuclear spins
in a polycrystalline adamantane sample, with $^{1}H$ nuclear spins
forming the environment. The solid lines represent the theoretical
magnetization signal calculated using the measured parameters. (b)
The attenuation factor $\mathcal{J}(\tau_{c},t)$, computed from the
experimental magnetization signal for each measurement time, which
allows for the estimation of $\tau_{c}$ as a function of time using
different approximations. The solid curves represent the theoretical
attenuation factor for the given parameters. Inset: The fitted spectral
density $G(\tau_{c},\omega)$, estimated from the experimental data
using noise spectroscopy. The spectral density confirms the Lorentzian
profile with parameters consistent with the measured values.}
\label{M-J-exp}
\end{figure}

Before estimating $\tau_{c}$, we first characterize the spin environment
by determining the noise spectral density $G(\tau_{c},\omega)$. This
step is essential because it provides a reference for later comparing
the estimated $\tau_{c}$ at different measurement times. Without
this initial characterization, it would be unclear whether discrepancies
in $\tau_{c}$ arise from the estimation process itself or from deviations
in the actual environmental noise spectrum. To obtain $G(\tau_{c},\omega)$,
we perform dynamical decoupling noise spectroscopy following the protocol
in Ref. \citep{Alvarez2011}. We measure the magnetization decay of
the $^{13}C$ qubit-probes using CPMG
sequences. The resulting spectral density, shown in the inset of Fig.
\ref{M-J-exp}, exhibits a Lorentzian shape with parameters $g=8.58$
kHz and $\tau_{c}=0.08$ ms, represented by the solid line.

Determining the spectral density requires data acquisition across
multiple time points, making it experimentally demanding due to its
non-parametric nature. In contrast, our critical-transition-based
approach suggests that optimal estimation of $\tau_{c}$ may be achievable
with far fewer measurements. However, this estimation depends on an
assumed spectral density model, as it involves parametric inference
rather than direct reconstruction. This reveals a key advantage of
optimal inference methods, especially near the critical transition
where the estimation process exhibits control-dependent abrupt changes.
Such behavior highlights the importance of information-theoretic tools
for efficient environmental characterization.

To estimate $\tau_{c}$ from experimental data, we reproduce the estimation
protocol introduced in Sec. \ref{sec:III_Estimation-of-the}. We measure
the magnetization decay of the qubit-probes under CPMG sequences with
a fixed number of pulses $N$ and varying the delay between pulses
to control the total measurement time. This procedure is illustrated
in Fig. \ref{M-J-exp}(a). We consider the cases $g\,\tau_{c}\,\sqrt{2N}=1.4$
(for $N=2$, green squares) and $g\,\tau_{c}\,\sqrt{2N}=9.7$ ($N=100$,
blue triangles). From these magnetization decay signals, we extract
the attenuation factor as a function of the probing time (Fig. \ref{M-J-exp}(b)).
Using this observable, we estimate $\tau_{c}$ for different measurement
times following the inference strategies described in Sec. \ref{sec:III_Estimation-of-the}.

\begin{figure}[h]
\includegraphics[width=1\columnwidth]{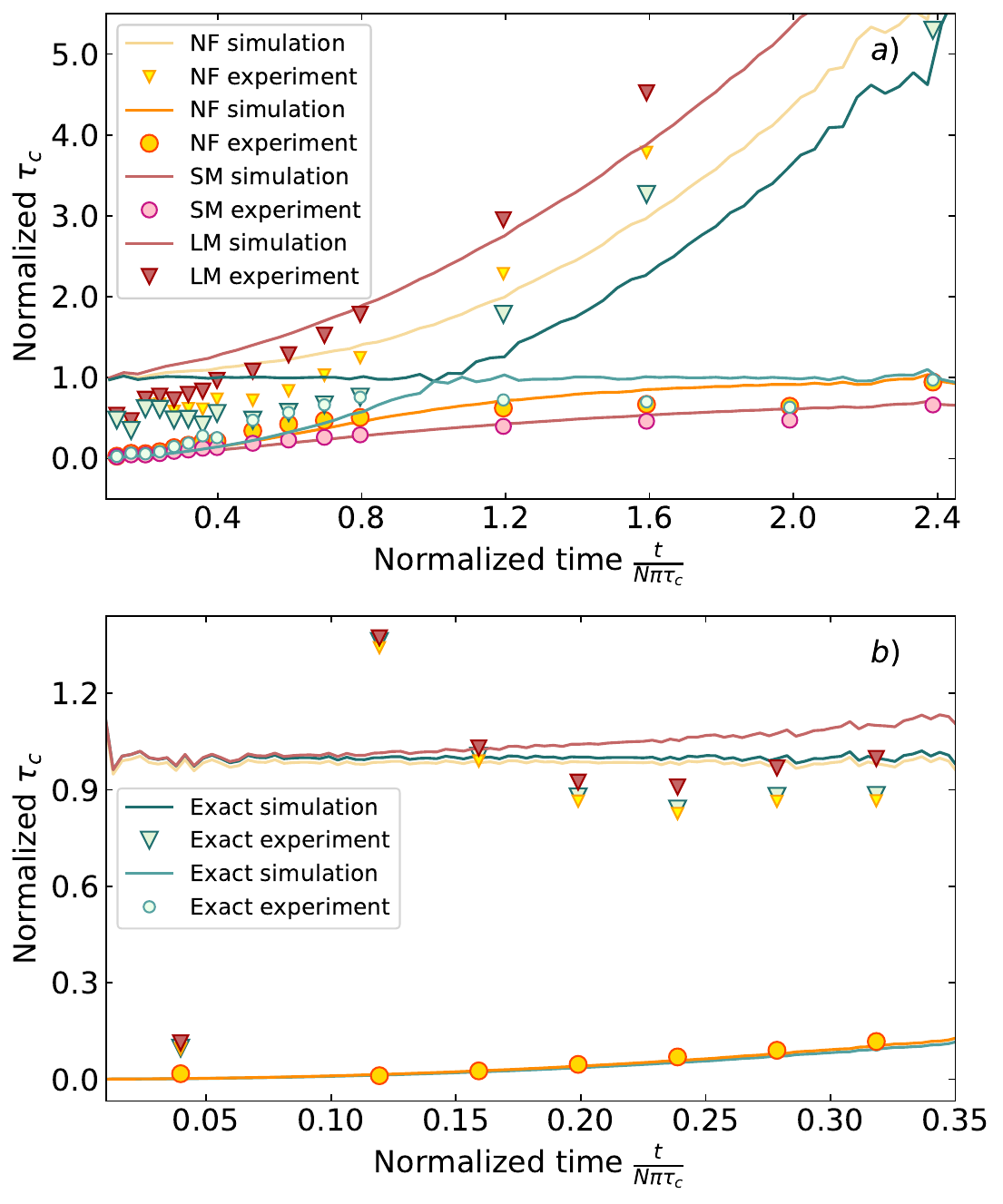}

\caption{Comparison between estimation using experimental data and simulations
for $g=8.58\,\text{kHz}$, $\tau_{c}=0.08\,\text{ms}$ as the true value, and (a) $N=2$
pulses $(\ensuremath{g\tau_{c}\sqrt{2N}=1.4})$ and (b) $N=100$
pulses $(\ensuremath{g\tau_{c}\sqrt{2N}=9.7})$. The estimated correlation
time $\tau_{c}$ is shown for different estimation strategies, normalized
by the true value, as a function of the measurement time. The data
includes both experimental results and simulations. In the simulations,
the true value of $\tau_{c}$ is set manually, while in the experiments,
it is estimated by fitting the spectral density.}
\label{experimental-1}
\end{figure}

Figure \ref{experimental-1} compares the estimated $\tau_{c}$ from
experimental data with the theoretical predictions obtained in the
previous section. Fig. \ref{experimental-1}(a) shows the case $g\,\tau_{c}\,\sqrt{2N}=1.4$. The experimental results exhibit similar trends
to the theoretical simulations, confirming the expected avoided-crossing
behavior. However, the crossing seems to occur at a lower measurement time
than predicted. Figure \ref{experimental-1}(b) presents the results for \( g\,\tau_{c}\,\sqrt{2N} = 9.7 \), which also follows the trend predicted by simulations. Note that, as in Fig. \ref{Fig3}, only one dynamical regime is accessed here.

In the SM regime, the experimental estimation recovers the expected
value of $\tau_{c}$. However, it is important to note that experimentally,
we do not have direct access to the true value. Instead, our reference
value is derived from the spectral density fitting (inset Fig. \ref{M-J-exp}(b)),
which may introduce discrepancies due to deviations from a perfect
Lorentzian functional form at high frequencies. These discrepancies are particularly noticeable in the LM regime, which probes higher frequencies, where the experimental estimation deviates more significantly from theoretical predictions.

For the case $N=2$ with $g\tau_{c}\sqrt{2N}=1.4$, both the LM and
SM regimes should, in principle, provide similar estimation accuracy
(see Fig. \ref{fig2_Criticality-schematic} inset (b) and Fig. \ref{Fig3}(g)). However, due to experimental constraints imposed by the sample,
the values of $g$ and $\tau_{c}$ are fixed, limiting our ability
to freely explore the critical condition. This constraint requires
us to use a low $N=2$ to satisfy $g\tau_{c}\sqrt{2N}=1.4$. Under
these conditions, the narrow-band filter approximation is valid when
the measurement time satisfies $t/\pi N>\tau_{c}$, making the SM
regime a better approximation in this experimental setup. This might
explain why the estimation process agrees more closely with the simulations
in Fig. \ref{experimental-1}.

The experimental results demonstrate a clear transition between different
regimes, consistent with simulations. This critical behavior manifests
as a sharp change in the estimation of $\tau_{c}$ around $t\approx\pi N\tau_{c}$,
which reflects a significant shift in the system\textquoteright s
dynamics. While the Cramer-Rao bound suggests that the information
at the critical point itself is minimal, the observation of this transition
provides important insight into the underlying system and its response
to the measurement time.

The critical transition serves as a diagnostic signal that guides
the estimation process, identifying regimes with distinct dynamical
behaviors. This enables more informed $\tau_{c}$ estimation, although capturing full system dynamics near criticality still requires
extensive measurements. While criticality does not directly reduce
the number of required measurements, it provides a valuable tool for
understanding the system's behavior and optimizing measurement strategies.

\section{Estimation error analysis\label{sec:Estimation-error-analysis}}

In this section, we analyze the relative error of the estimation obtained
from our simulations and experiments, comparing it with theoretical
predictions. We first examine the estimation results from the simulated
experiments described in Sec. \ref{sec:Estimating-the-memory-time}.
The third row of Fig. \ref{Fig3} displays the predicted relative
error per measurement $\varepsilon_{R}$ as solid black lines, corresponding
to the insets of Fig. \ref{fig2_Criticality-schematic}. The green
symbols in Fig. \ref{Fig3} (middle rows) represent the relative error
obtained from estimations using the exact expression for the attenuation
factor.

To calculate this error as a function of time, we follow these steps:
Compute the exact correlation time for the $N_{\text{rep}}$ sets of $N_{m}$ measurements
using the corresponding average magnetization for each measurement
time. For each measurement time, determine the two possible solutions
for $\tau_{c}$. Calculate the standard deviation of the $N_{\text{rep}}$
values for each solution relative to the true value of $\tau_{c}$. We present the error per measurement, obtained by multiplying the estimated error by $\sqrt{N_{m}}$, since the error scales inversely with the square root of the number of measurements \citep{Caves_1994_fisher,Paris2009_QUANTUM-ESTIMATION}.

The time evolution of this standard deviation is shown in Fig. \ref{Fig3}(g)-(i)
for different parameter sets, indicated by green symbols. When comparing
these results to the theoretical behavior described in Eq. (\ref{eq:error_QFI}),
we find that they match the predictions on either side of the critical
point. However, the divergence is not reproduced, which is expected
since the control filters are not ideal delta functions.

Specifically:
\begin{itemize}
\item For (i) $g\,\tau_{c}\,\sqrt{2N}=1.4$, near the transition between
regimes (Fig. \ref{Fig3}(g)), the relative error is observed for
the two possible solutions. Criticality manifests as a crossing between
the errors associated with each solution. The theoretical bound for
the error assumes an unbiased measurement. On either side of the critical
point, one of the solutions becomes biased, resulting in an error
significantly larger than the theoretical prediction.
\item For (ii) $g\,\tau_{c}\,\sqrt{2N}=9.7$, corresponding to the LM regime
(Fig. \ref{Fig3}(h)), the relative error is examined. The solution
aligned with the LM approximation matches the theoretical error well.
However, the other solution, representing a biased measurement, deviates
from the theoretical prediction.
\item For (iii) $g\,\tau_{c}\,\sqrt{2N}=0.13$, in the SM regime (Fig. \ref{Fig3}(i)),
the relative error is consistent with the theoretical prediction for
times greater than the critical point. Near the critical point, the
transition between the LM and SM regimes is evident in the relative
error. These simulations demonstrate that the critical point can be
inferred from the relative error of $\tau_{c}$.
\end{itemize}

\begin{figure}[th]
\includegraphics[width=1\columnwidth]{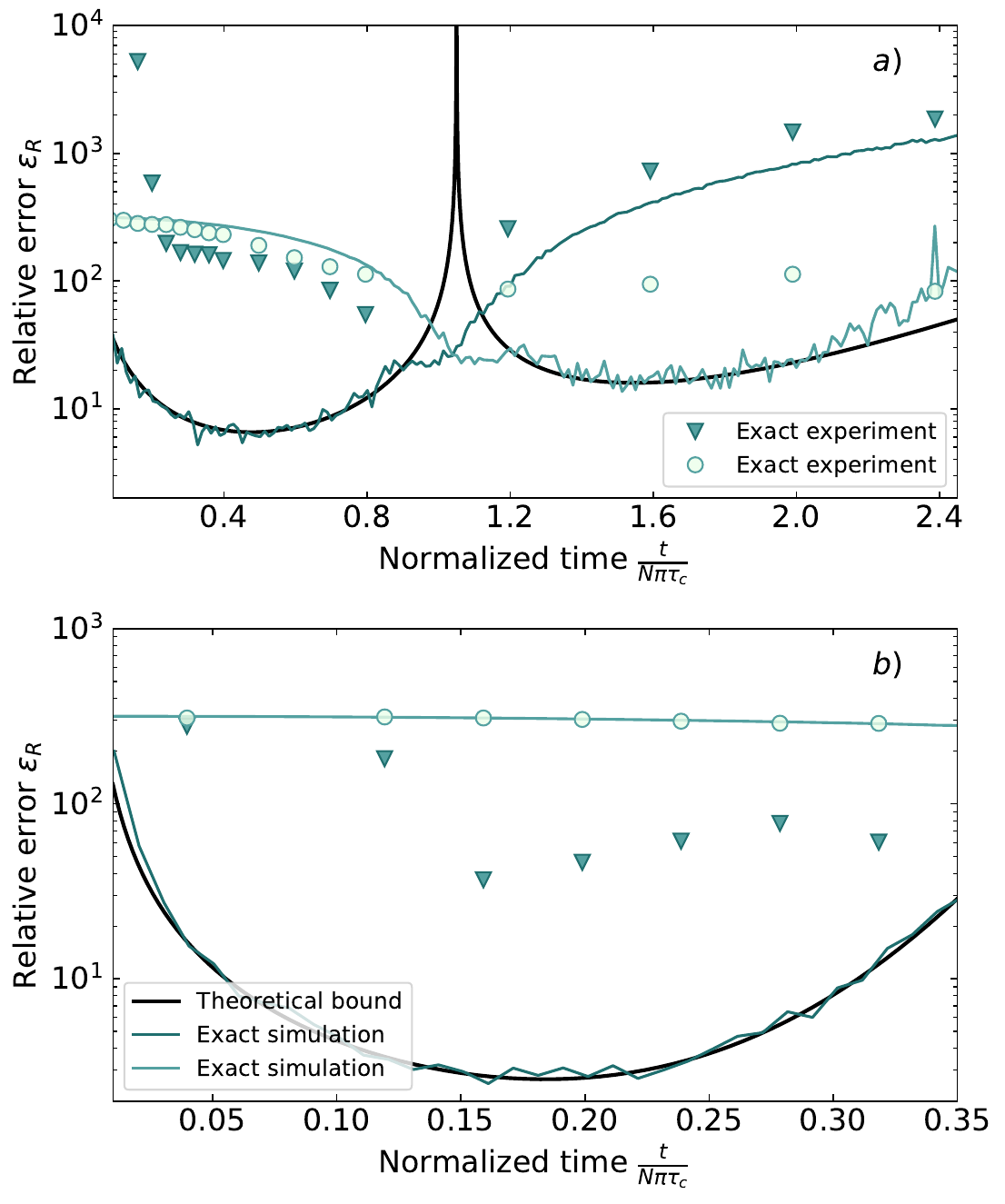}

\caption{Estimated relative error $\varepsilon_{R}$ per measurement in the
estimation of $\tau_{c}$. Comparison between experimental data and
simulations for $g=8.58\,$kHz, $\tau_{c}=0.08\,$ms as the true value and (a) $N=2$ pulses
($g\,\tau_{c}\,\sqrt{2N}=1.4$) and (b) $N=100$ pulses
($g\,\tau_{c}\,\sqrt{2N}=9.7$). $\varepsilon_{R}$ is the numerical
estimation of $\tau_{c}$ using the analytical attenuation factor
$\mathcal{J}$, comparing experimental results with simulations. The error is obtained by the mean distance of each of the 2 solutions to the real value. The
black continuous curve represents the theoretical relative error considering
the attenuation factor for the given parameters. $\varepsilon_{R}$
is computed to assess the precision of the estimation and highlights
the critical behavior observed in the transition region.}
\label{experimental-error}
\end{figure}

Figure \ref{experimental-error} shows the relative error obtained
from experimental estimations displayed in Fig. \ref{experimental-1}
for (a) $g\,\tau_{c}\,\sqrt{2N}=1.4$ and (b) $g\,\tau_{c}\,\sqrt{2N}=9.7$. This experimental error is compared with the theoretical predictions
and the results from simulated data shown in Fig. \ref{Fig3}(g)-(h).

In the experiments, instead of a single qubit sensor, we work with a macroscopic sample containing an ensemble of independent spins that act as qubit sensors, each interacting with its own environment. As a result, each measurement inherently provides an average signal from all the qubit sensors simultaneously. This is analogous to repeating the experiment multiple times using a single qubit sensor, as assumed in the simulations. However, in the experiment, it is not possible to precisely determine the effective number of measurements in advance, as it depends on the number of spins effectively detected by the NMR coil. For presentation purposes, we obtained an approximate experimental error per measurement by multiplying the experimental relative error by $\sqrt{N_{m}}=\sqrt{10^{5}}$, which was the number used in the simulated data. This allowed for a good alignment of the minimum value with the theoretical predictions shown in Fig. \ref{experimental-1}.

In Fig. \ref{experimental-error}(a), corresponding to \( g\,\tau_{c}\,\sqrt{2N} = 1.4 \), and Fig. \ref{experimental-1}(a), the relative error from the experiments qualitatively follows the predicted curves in the short-memory (SM) regime. However, in the long-memory (LM) regime, the error for one of the solutions deviates from the simulated predictions. This discrepancy is expected, as the estimated value differs from the true value in this regime (Fig. \ref{experimental-1}(a)). Nevertheless, despite the experimental data not perfectly matching the theoretical model, clear evidence of a critical transition is observed through noticeable changes in the relative error behavior near the critical point.

In Fig. \ref{experimental-error}(b), corresponding to \( g\,\tau_{c}\,\sqrt{2N} = 9.7 \), and Fig. \ref{experimental-1}(b), the experimental relative error again qualitatively aligns with the theoretical predictions. Significant deviations arise in regions where the estimated values deviate from the expected true values (Fig. \ref{experimental-1}(b)). As discussed earlier, these discrepancies in the LM regime are likely due to the experimental spectral density not perfectly matching a Lorentzian profile, particularly at the high frequencies probed in this regime (inset Fig. \ref{M-J-exp}(b)), or due to the filter function not being sufficiently narrow.
 
\section{Conclusions}

In this work, we have demonstrated the emergence of critical behavior in the estimation of environmental parameters using a dynamically controlled quantum sensor. By analyzing the estimation of the environment's correlation time across the transition between short-memory (SM) and long-memory (LM) dynamical regimes, we identified a sharp crossover point where the estimation becomes highly sensitive. At this critical point, the uncertainty in the inferred correlation time---quantified via an information-theoretic metric---diverges, reflecting a vanishing value for the Fisher information. Both experimental measurements and numerical simulations corroborate the theoretical prediction of this criticality in parameter estimation.

This criticality arises from the transition between two distinct solutions
for the correlation time, each corresponding to a different dynamical
regime. At the transition point, a sudden jump between these solutions
occurs, leading to a noticeable increase in the estimation error.
This behavior highlights the sensitivity of the probe near criticality,
emphasizing the importance of careful probe control optimization in this
region. Our results show that while the LM regime offers high precision
for short measurement times, the SM regime provides advantages for
long-time settings.

The relative error analysis highlights the practical significance of the observed critical behavior in estimating the environmental correlation time $\tau_c$. Near the critical point, the estimation exhibits a characteristic non-monotonic behavior, with a sharp increase in the relative error that signals the presence of two competing solutions---each corresponding to a different dynamical regime. This behavior resembles an avoided crossing and serves as a diagnostic tool: By analyzing its landscape, one can pinpoint the transition and distinguish between dynamical regimes, and therefore one can select the one in which the estimation of $\tau_c$ is more accurate and reliable. Thus, criticality not only reveals fundamental aspects of the probe-environment interaction and system dynamics, but also informs the design of more effective parameter estimation strategies.

Furthermore, these insights lay the groundwork for developing adaptive
measurement protocols. In such protocols, the quantum sensor can be
dynamically controlled to exploit the critical point, where the Fisher
information vanishes and the sensor's sensitivity is expected to fluctuate
dramatically due to the lack of information at the transition. This
behavior can provide an alternative approach for parameter estimation,
where measurement times and control parameters are carefully adjusted
based on the system\textquoteright s behavior near the critical point,
optimizing the estimation process despite the information limitations.

Future work could extend this methodology to environments whose spectral densities deviate from the ideal Lorentzian profile, enabling the exploration of criticality-driven sensing in more complex and realistic systems. By understanding and leveraging the emergence of critical estimation behavior in such settings, it may be possible to control quantum sensors to probe dynamic behaviors in applications ranging from biological and chemical detection to the study of complex quantum materials. In particular, materials exhibiting spin or charge dynamics that give rise to correlated fluctuations such as quantum magnets, low-dimensional conductors, and disordered spin networks, represent systems of growing interest for quantum sensing applications. These environments provide exciting platforms for testing and expanding criticality-driven sensing approaches. Insights gained from these studies could also inform the design of quantum control protocols tailored for robustness and efficiency in practical sensing scenarios.

Acknowledgments.--- This work was supported by CNEA; CONICET; Fundacion
Balseiro, ANPCyT-FONCyT PICT-2017-3156, PICT-2017-3699, PICT-2018-4333,
PICT-2021-GRF-TI-00134, PICT-2021-I-A-00070; PIP-CONICET (11220170100486CO,
11220220100531CO); UNCUYO SIIP Tipo I 2019-C028, 2022-C002, 2022-C030;
PIBAA 2022-2023; Instituto Balseiro; Collaboration programs between
the MINCyT (Argentina) and MAECI (Italy) and MOST (Israel).
\bibliographystyle{apsrev}
\bibliography{bibliography,references}

\end{document}